# Agentic AI for Cybersecurity: A Meta-Cognitive Architecture for Governable Autonomy

**Authors**: Andrei Kojukhov and Arkady Bovshover

## Abstract

Cybersecurity decision-making increasingly occurs in environments characterized by uncertainty, partial observability, and adversarial manipulation, where heterogeneous signals from multiple sources are often incomplete, ambiguous, or conflicting. Traditional Security Orchestration, Automation, and Response (SOAR) systems rely on deterministic pipelines and threshold-based triggers, limiting their ability to support reliable decision-making under such conditions.

This paper proposes a probabilistic, agentic framework for cybersecurity orchestration that models decision-making as a **meta-cognitive process**. The framework decomposes cybersecurity functions into interacting agents responsible for detection, hypothesis formation, contextualization, explanation, and governance, coordinated through a meta-cognitive judgement mechanism. This mechanism evaluates uncertainty, agent disagreement, and operational constraints to determine decision readiness, enabling adaptive strategies including automated action, escalation, deferral, and evidence refinement.

Empirical evaluation on benchmark datasets (CICIDS2017 and NSL-KDD), augmented with adversarial and uncertain conditions, demonstrates that the proposed approach improves robustness and decision quality compared to deterministic and single-agent baselines. The framework achieves higher accuracy under noise, reduces false positive rates, and produces better-calibrated confidence estimates, while enabling more adaptive and context-aware decision behavior.

By explicitly modeling meta-cognitive processes—monitoring, evaluation, control, and reflection—the proposed approach reframes cybersecurity as an instance of **AI-mediated cognitive problem solving**, supporting accountable autonomy and more effective human–AI collaboration in adversarial environments.

## 1. Introduction

Cybersecurity has undergone a profound transformation over the past decade. Once dominated by signature-based detection and rule-driven automation, modern cyber defence increasingly relies on machine learning, large-scale data analytics, and automated response mechanisms to cope with the growing volume and velocity of security events. These advances have significantly improved detection coverage and operational efficiency.

However, they have also revealed structural limitations in how cybersecurity systems are conceptualized and designed.

Most contemporary AI-enabled cybersecurity platforms are organized as **pipelines [1,2]**. Telemetry is ingested, processed by detection models, filtered into alerts, and optionally translated into automated responses. This architectural paradigm implicitly treats AI as a collection of tools embedded within a largely linear workflow. Even when multiple models are employed, coordination logic remains centralized and opaque, with limited support for reasoning under uncertainty, explanation of decisions, or meaningful human–AI collaboration [1,7,8].

At the same time, the operational reality of cybersecurity is changing. Recent industry analyses document the emergence of **agentic AI systems** capable of planning, reasoning, and acting autonomously across both offensive and defensive cyber operations. Attackers increasingly employ AI to automate reconnaissance, generate adaptive phishing campaigns, and dynamically modify attack strategies. Defenders, in turn, deploy AI-driven systems that operate continuously, make real-time decisions, and interact directly with production environments. In this context, cybersecurity is no longer a problem of isolated detections, but one of **coordinated decision-making under adversarial uncertainty**.

Recent work surveys the evolving role of agentic AI in cybersecurity, tracing a shift from single-step generative models to multi-agent systems and semi-autonomous pipelines that require continuous reasoning, tool use, and iterative decision cycles [17, 18]. This evolution motivates the need for architectures that can not only coordinate agentic behavior but also govern decision autonomy under uncertainty.

Academic surveys of artificial intelligence in cybersecurity provide a complementary perspective. They document substantial progress in applying machine learning and deep learning to tasks such as intrusion detection, malware classification, and anomaly detection [1,2]. However, these surveys also reveal an implicit assumption that AI systems operate as bounded, task-specific components. Explainability, governance, and trust are typically treated as secondary concerns or future challenges [7,8,10], rather than as integral elements of system design.

Recent comprehensive surveys of artificial intelligence in cybersecurity provide valuable overviews of machine learning, deep learning, and natural language processing techniques applied across detection, response, and threat intelligence tasks [16]. However, such accounts largely assume pipeline-oriented architectures and do not explicitly address how autonomous decisions are coordinated, justified, and governed under adversarial uncertainty.

This disconnect between **model-centric academic framings** and **agentic operational realities** points to a deeper conceptual gap. We argue that cybersecurity should be understood not merely as a technical pipeline, but as a **distributed cognitive process [4,5,6]** involving multiple artificial and human agents. From this perspective, orchestration is not simply automation, but coordination, negotiation, explanation, and accountability.

The central contribution of this paper is a theoretical re-framing of cybersecurity orchestration as an **agentic, multi-agent cognitive system**, enabled by generative AI and grounded in principles of responsible and explainable AI. By making this structure explicit, we aim to provide a foundation for designing cybersecurity systems that are more adaptive, interpretable, and governable in the face of growing complexity.

## 1.1 Positioning Relative to SOAR Architectures

Security Orchestration, Automation, and Response (SOAR) platforms play a central role in contemporary security operations by coordinating tools, alerts, and predefined response workflows [12]. These systems have significantly improved operational efficiency by automating repetitive tasks and enforcing consistent playbooks. However, SOAR architectures remain fundamentally action-centric: they orchestrate responses rather than the reasoning processes [4,5] that precede and justify those responses.

The framework proposed in this paper is not an extension or replacement of SOAR, nor does it seek to automate additional security tasks. Instead, it operates at a different conceptual layer. We focus on orchestrating epistemic functions—such as interpretation, hypothesis formation, contextual integration, explanation, and governance [7,8,10] —that are largely implicit or externalized in existing SOAR systems. In practice, a SOAR platform may serve as an execution substrate within our framework, but it does not provide the cognitive coordination required for decision-making under uncertainty.

Similarly, recent integrations of large language models into SOC workflows typically position generative AI as an interface layer for summarization or analyst interaction. In contrast, we position generative AI as a coordination substrate that enables semantic negotiation [7,9,11] and alignment across heterogeneous agents and human analysts. This distinction reflects a shift from automating actions to governing autonomy, which becomes increasingly critical as AI systems assume more active roles in cyber defence.

## 1.2 Positioning Relative to Multi-Agent Systems Theory

The proposed framework draws inspiration from multi-agent systems (MAS) research but differs from classical MAS formulations in both objective and scope. Traditional MAS theory focuses on coordination, negotiation, and optimization [3] among autonomous agents operating under predefined goals and utility functions. In contrast, cybersecurity decision-making is characterized by ambiguous objectives, adversarial uncertainty [1], and strong accountability constraints, where the correctness of an action cannot be evaluated solely in terms of optimality or convergence.

Rather than treating autonomy as a property to be maximized, we treat it as a governable and revisable attribute. Explainability, policy compliance, and human oversight [7,8,10,12] are therefore modelled as first-class cognitive functions, not as external constraints applied after agent coordination. This shifts the role of agent interaction from optimizing behavior to justifying decisions under uncertainty, a distinction that is not central in classical MAS frameworks.

Accordingly, our contribution is not the application of existing MAS algorithms to cybersecurity tasks, but the articulation of an architectural and cognitive perspective in which agent interaction serves accountable decision-making rather than purely instrumental coordination.

This perspective aligns more closely with theories of distributed cognition and socio-technical systems than with classical MAS optimization paradigms.

# 2. Theoretical Foundations and Meta-Cognitive Framework

## 2.1 From Linear Pipelines to Distributed Cognition

Traditional cybersecurity architectures are rooted in a pipeline metaphor [1,2]: data flows through predefined stages, each performing a narrowly defined function. This metaphor assumes that decision-making can be decomposed into independent steps and that uncertainty can be resolved locally at each stage. While effective for deterministic or well-understood threats, this assumption breaks down in adversarial environments where signals are ambiguous, incomplete, and often contradictory.

We propose an alternative theoretical framing: **cybersecurity as a distributed cognitive system**. In this view, security operations resemble human cognitive activity more than industrial automation. Observations must be interpreted, hypotheses formed and revised, context integrated, decisions explained, and actions justified within organizational and regulatory constraints. These processes are inherently interactive, iterative, and social rather than linear.

From a cognitive perspective, these processes form a continuous loop of monitoring, interpretation, evaluation, and action. Unlike linear pipelines, decision-making emerges from iterative refinement of hypotheses under uncertainty, where new evidence may reinforce, contradict, or reframe prior conclusions. This dynamic process is central to cybersecurity operations in adversarial environments.

## 2.2 Agentic Decomposition of Cybersecurity Functions

An agentic perspective enables a principled decomposition of cybersecurity into specialized cognitive roles, implemented as interacting agents:

- **Detection** identifies and surfaces weak or ambiguous signals from heterogeneous data sources.
- **Hypothesis Formation** supports abductive reasoning by generating and revising competing explanations under uncertainty.
- **Contextualization** integrates organizational, temporal, and situational knowledge to inform interpretation and decision-making.
- **Explanation** translates machine-generated reasoning into forms that are interpretable, inspectable, and actionable by human analysts.
- **Governance** enforces policy, ethical, and regulatory constraints on permissible actions and system behavior.
- **Meta-cognitive judgement** operates as defined in Definition 1, integrating uncertainty signals, explanation adequacy, and governance constraints to assess decision readiness and regulate action authority.

**Definition 1 (Meta-Cognitive Judgement).**

Meta-cognitive judgement is the system-level capacity of an AI-assisted cybersecurity architecture to determine whether, when, and under what authority conditions an action should be executed, given uncertainty, agent disagreement, operational risk, and accountability constraints. Unlike predictive accuracy or control logic, meta-cognitive judgement governs decision readiness: it evaluates the legitimacy, proportionality, and autonomy of actions before execution, including the need for deferral, escalation, or human oversight.

Operationally, meta-cognitive judgement can be understood as a regulatory loop that integrates multiple cognitive processes: monitoring of agent outputs, evaluation of uncertainty and disagreement, control over action selection, and reflection through feedback and adaptation. This loop governs not only what the system predicts, but whether its predictions are sufficiently reliable and justified to support action.

Crucially, no single agent possesses global authority or complete knowledge. Security decisions emerge through **coordination and negotiation**, rather than centralized control. This mirrors human team-based decision-making and provides a natural foundation for accountable autonomy.

## 2.3 Generative AI as a Semantic Coordination Substrate

Generative AI plays a pivotal role in enabling agentic orchestration. Unlike traditional predictive models, generative systems support semantic reasoning, dialogue, and adaptive explanation. These capabilities allow agents to exchange hypotheses, express uncertainty, align interpretations, and interact naturally with human analysts.

From a theoretical perspective, generative AI functions as a **cognitive coordination substrate**, enabling distributed reasoning across heterogeneous agents.

## 2.4 Explainability and Governance as Cognitive Functions

A defining aspect of the proposed framework is the elevation of explainability and governance from peripheral concerns to **first-class cognitive functions**. Explanation is treated not as post-hoc reporting, but as an active constraint [7,8] on decision-making. Governance similarly operates continuously, mediating between automated recommendations and permissible actions [10,11,12].

This approach aligns autonomy with accountability and reflects emerging responsible AI principles that emphasize transparency, traceability, and meaningful human control.

Meta-cognitive judgement is operationalized through the interaction between explanation and governance, serving as the mechanism through which system recommendations become procedurally legitimate decisions rather than opaque outputs.

## 2.5 Meta-Cognition and Meta-AI as Foundations of Judgement

The meta-cognitive judgement function in the proposed framework is explicitly meta-cognitive, operating at a level distinct from domain-level threat reasoning [13,14]. By regulating decision readiness rather than predictive accuracy, meta-cognitive judgement supports accountable autonomy consistent with responsible AI principles [14,15]. This meta-level role is instantiated architecturally through meta-cognitive judgement agents, as discussed in Section 3, where its operational implications are specified.

From a Meta-AI perspective, meta-cognitive judgement operates as a regulatory layer that governs when and how autonomous decisions may be taken. In this sense, meta-cognition enables proportional autonomy by dynamically calibrating the system's level of independence based on epistemic conditions. In this sense, meta-cognitive judgement operationalizes core meta-cognitive processes—monitoring, evaluation, control, and reflection—within a cybersecurity context. Monitoring corresponds to tracking distributed signals and agent outputs; evaluation assesses uncertainty and coherence; control governs action authority; and reflection enables adaptation based on outcomes. This explicit mapping clarifies how cognitive processes are embedded within the system architecture.

By embedding meta-cognitive judgement directly into the architecture, the framework supports reflective autonomy rather than reactive automation, aligning AI decision-making with principles of accountability, transparency, and meaningful human control.

Related educational research on Meta-AI skills and human–AI collaboration has similarly emphasized meta-cognitive monitoring and regulation as prerequisites for effective interaction with generative AI systems.

Conceptual work on metacognitive AI foregrounds the value of self-monitoring and adaptive reasoning for managing uncertainty and enhancing accountability in autonomous systems [21], aligning with our use of meta-cognitive judgement as a structural oversight mechanism.

Taken together, these elements establish cybersecurity decision-making as a meta-cognitive process in which actions are not triggered directly by detections but are mediated by the system's assessment of its own knowledge, uncertainty, and decision readiness. This perspective provides the conceptual foundation for the agentic orchestration framework described in the following section.

# 3. Architecture and Framework: Agentic Cybersecurity Orchestration

## 3.1 Architectural Overview

Building on the theoretical framing, we propose an **Agentic Cybersecurity Orchestration Framework** that operationalizes cybersecurity as a multi-agent cognitive system. The architecture consists of five interacting agent classes coordinated through an orchestration layer powered by generative AI, with humans integrated as active participants.

Unlike traditional SOC architectures, which centralize logic in SIEMs or automation engines, this framework distributes reasoning across agents with distinct epistemic responsibilities, increasing resilience and transparency.

## 3.2 Agent Roles

- **Detection Agents** ingest raw telemetry and surface probabilistic signals.
- **Hypothesis Agents** generate and manage competing explanations.
- **Context Agents** integrate business, temporal, and threat intelligence context.
- **Explainability Agents** ensure decisions remain interpretable and defensible.
- **Meta-cognitive judgement Agents** operate at the meta-level of the architecture, assessing the adequacy and legitimacy of system decisions. Drawing on principles of meta-cognition, these agents monitor uncertainty, confidence alignment across contributing agents, completeness of contextualization, and the sufficiency of explanations relative to governance constraints. Rather than optimizing outcomes or enforcing policy, judgement agents assess decision readiness: whether available evidence, interpretations, and constraints justify autonomous action, require human escalation, or warrant continued deliberation. In this role, judgement functions as a regulatory mechanism that dynamically calibrates system autonomy under adversarial uncertainty.

- **Governance Agents** encode policy, compliance, and ethical constraints.

Each agent is explicitly designed to be **fallible and revisable**, supporting proportional and accountable decision-making.

## 3.3 Orchestration and Human Oversight

The orchestration layer enables communication, negotiation, and coordination among agents. Generative AI facilitates semantic alignment and adaptive explanation, while providing a focal point for logging, audit, and accountability.

Human analysts interact with the system at multiple levels, from hypothesis evaluation to policy override, enabling **graduated autonomy** consistent with regulatory expectations.

Meta-cognitive judgement provides a natural locus for human oversight, enabling analysts to interrogate not only system outputs, but also the procedural rationale and criteria by which decision readiness and autonomy thresholds are evaluated and revised.

## 3.4 Conceptual Architecture

Below Figure 1 describes the conceptual architecture of agentic cybersecurity orchestration. The framework is organized as a network of interacting agents coordinated by a generative orchestration layer, with humans integrated as active participants. Domain-level agents support detection, hypothesis generation, contextual interpretation, explanation, and governance, while a meta-cognitive judgement function operates at a higher level to evaluate decision readiness and regulate system autonomy. This meta-level control enables reflective, accountable decision-making by mediating between competing interpretations, uncertainty, and policy constraints. The figure contrasts traditional linear pipelines with cyclic, negotiated, and reflective decision flows characteristic of agentic cybersecurity systems.

For example, in a login anomaly scenario, detection agents identify unusual access patterns, hypothesis agents generate competing explanations (e.g., benign travel versus account compromise), context agents incorporate user behavior history, and meta-cognitive judgement evaluates uncertainty and disagreement before determining whether to trigger automated response or escalate to a human analyst.

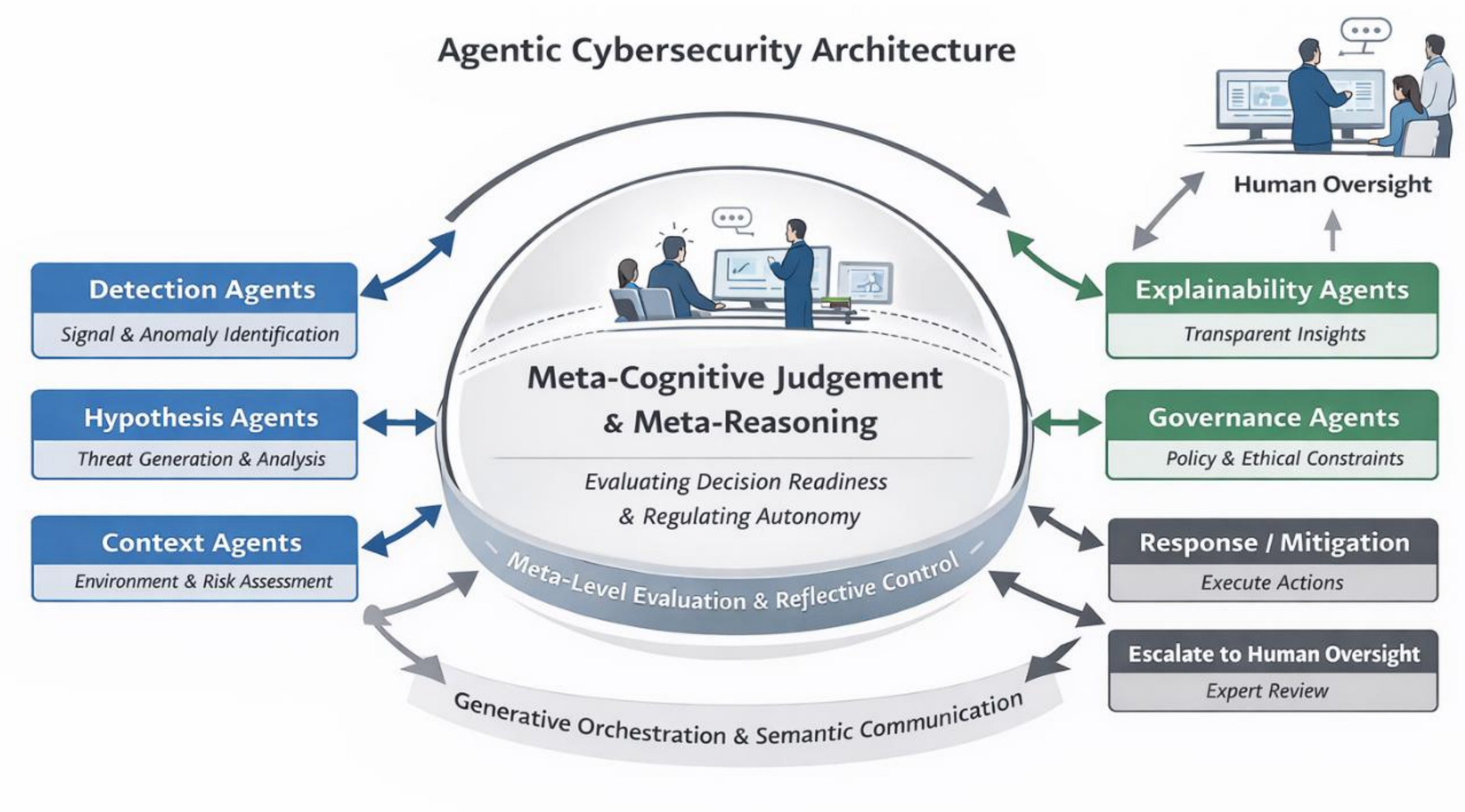

**Figure 1:** Conceptual Architecture of Agentic Cybersecurity Orchestration

Crucially, control does not flow linearly through the architecture; authority is conditionally granted by the meta-cognitive judgement function.

# 4. Related Work

## 4.1 AI in Cybersecurity: Model-Centric Foundations

Extensive prior research has explored the application of machine learning and deep learning to cybersecurity tasks, including intrusion detection, malware classification, phishing detection, and anomaly detection [1,2]. Survey papers systematically organize this literature by algorithm, dataset, and performance metric.

This body of work has produced substantial technical advances, but largely adopts a **model-centric, pipeline-oriented framing**.

Agentic AI systems introduce unique security challenges — including dynamic environmental interactions, adversarial exploitation of autonomy, and emergent behaviors unaccounted for in traditional frameworks — necessitating dedicated lifecycle-aware security approaches [19]. Complementary architectural proposals such as sentinel-based oversight highlight the value of distributed meta-control within multi-agent ecosystems [20].

## 4.2 Limitations of Existing Framings

In the reviewed literature, AI systems are typically treated as bounded components whose outputs are consumed by external decision logic. Coordination, uncertainty management, explanation, and governance are acknowledged but not structurally integrated [7,8,10].

As a result, explainability and responsible AI concerns are positioned as auxiliary rather than foundational.

### 4.3 Beyond Pipelines: Toward Agentic Perspectives

Emerging work on autonomous agents and reinforcement learning in cybersecurity begins to address adaptive decision-making but often focuses on narrow optimization problems and overlooks governance and accountability.

The present work advances this trajectory by proposing a **unifying agentic orchestration framework** that integrates technical performance with cognitive, organizational, and regulatory considerations.

### 4.4 Positioning and Policy Alignment

This paper positions cybersecurity autonomy as a **governable property**, not a technical inevitability. By embedding explainability, governance, and human oversight directly into the architecture, the proposed framework aligns with emerging responsible AI principles and policy frameworks.

The agentic orchestration perspective thus offers a foundation for designing cybersecurity systems that are not only effective, but also transparent, accountable, and aligned with societal and regulatory expectations [10,12].

## 5. Empirical Evaluation of Agentic Cybersecurity Orchestration

To complement the conceptual and architectural contribution, we evaluate the proposed framework using a combination of benchmark cybersecurity datasets and controlled simulation scenarios. The objective is to assess how agentic orchestration with meta-cognitive judgement improves decision-making under uncertainty compared to traditional pipeline-based approaches.

### 5.1 Evaluation Objectives

The evaluation focuses on system-level properties aligned with cognitive decision-making:

- robustness under noisy and adversarial conditions
- handling of conflicting and incomplete evidence
- calibration of confidence and decision readiness
- adaptive decision strategies (action, escalation, deferral)

Rather than optimizing only detection accuracy, we evaluate the system’s ability to **support reliable and accountable decisions** under uncertainty.

### 5.2 Datasets and Scenario Design

**Benchmark Datasets**

We use two widely adopted cybersecurity datasets:

- **CICIDS2017**: provides realistic network traffic including benign activity and diverse attack types (DDoS, brute force, infiltration).

- **NSL-KDD**: enables comparison across classical intrusion detection benchmarks with reduced redundancy.

These datasets are used to simulate heterogeneous agent inputs representing different detection modalities.

**Scenario Augmentation**

To approximate real-world Security Operations Center (SOC) conditions, we augment the datasets with controlled perturbations:

- **Missing signals**: randomly removing agent inputs
- **Conflicting evidence**: injecting contradictory outputs across agents
- **Adversarial noise**: perturbing features to simulate evasion
- **Temporal inconsistency**: introducing delayed or stale observations

This enables evaluation of decision-making under **uncertainty and disagreement**, rather than clean classification settings.

## 5.3 Agent-Based System Configuration

We simulate heterogeneous agents corresponding to real cybersecurity components:

- Network anomaly detection agent
- Endpoint behavior analysis agent
- User behavior analytics (UBA) agent
- Threat intelligence correlation agent
- Rule-based detection agent

Each agent produces:

- a probabilistic belief over hypotheses
- a confidence score reflecting local uncertainty

Agents are intentionally designed to be:

- partially observable
- heterogeneous in accuracy
- occasionally conflicting

## 5.4 Compared Approaches

We compare three system configurations:

1. **Deterministic SOAR Pipeline**
    - rule-based orchestration
    - threshold-triggered actions
2. **Single-Agent ML System**
    - centralized classifier
    - no explicit uncertainty integration
3. **Proposed Agentic Orchestration Framework**
    - multi-agent belief aggregation

- meta-cognitive decision function
- adaptive decision strategies

## 5.5 Evaluation Metrics

We evaluate both predictive and cognitive decision properties:

**Detection Metrics**

- accuracy
- precision / recall / F1

**Decision Metrics**

- correct action rate
- false action rate

**Cognitive Metrics**

- confidence calibration error
- entropy-based uncertainty behavior

**Operational Metrics**

- escalation rate
- decision latency
- refinement frequency

## 5.6 Results

The following results empirically validate the proposed meta-cognitive framework, demonstrating how monitoring, evaluation, control, and reflection improve decision-making under uncertainty.

**Detection Performance under Noise**

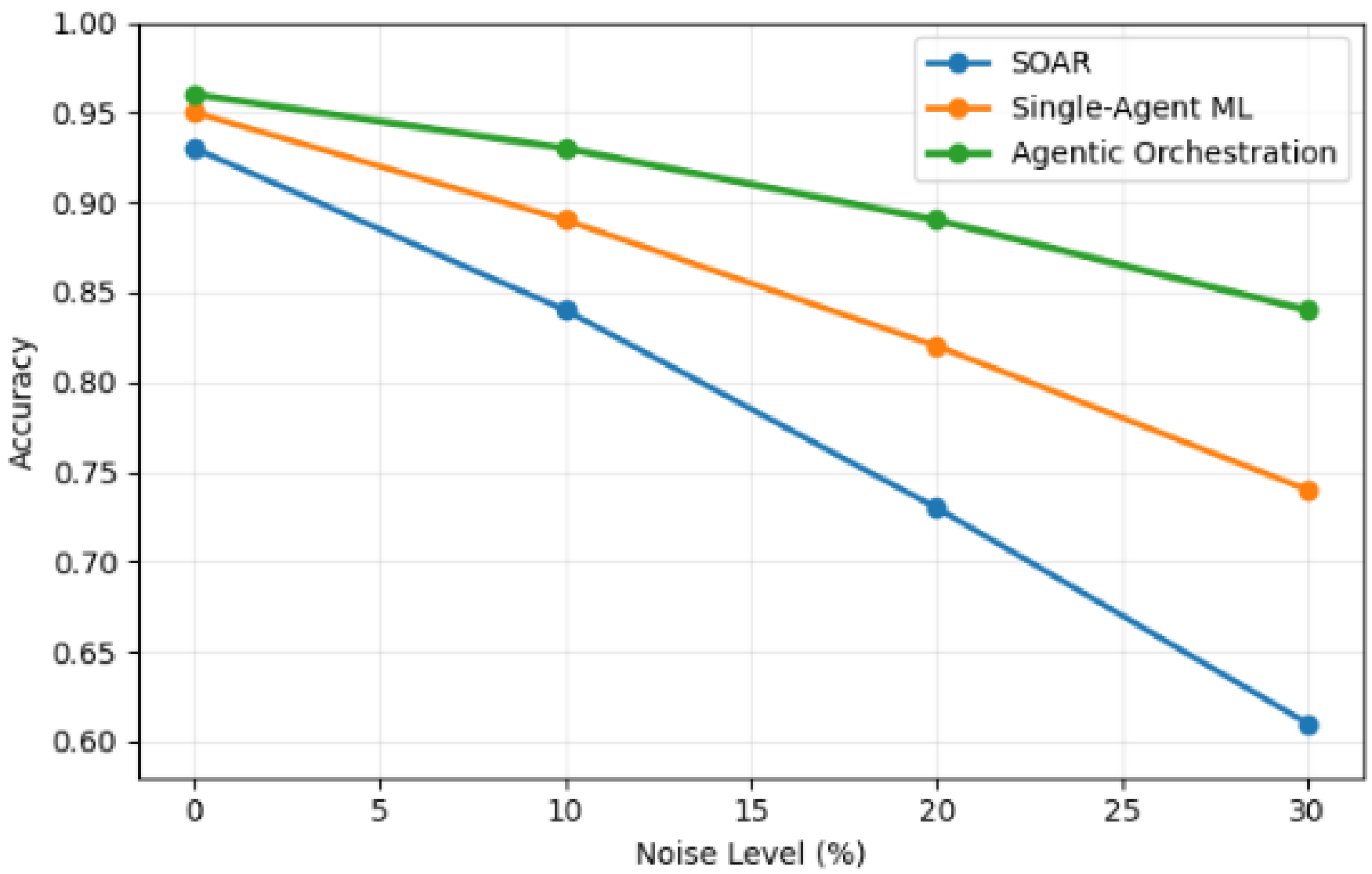

Figure 1 shows detection accuracy under increasing noise levels (0%–30%). While all methods perform similarly under clean conditions, the deterministic SOAR pipeline degrades sharply as noise increases, dropping from approximately 0.93 to 0.61 accuracy. The single-agent model shows moderate robustness, maintaining accuracy around 0.74 at the highest noise level.

In contrast, the proposed framework maintains accuracy above 0.84 under high noise conditions, demonstrating more stable performance. This indicates that multi-agent aggregation improves resilience to incomplete and uncertain observations.

**False Positive Reduction**

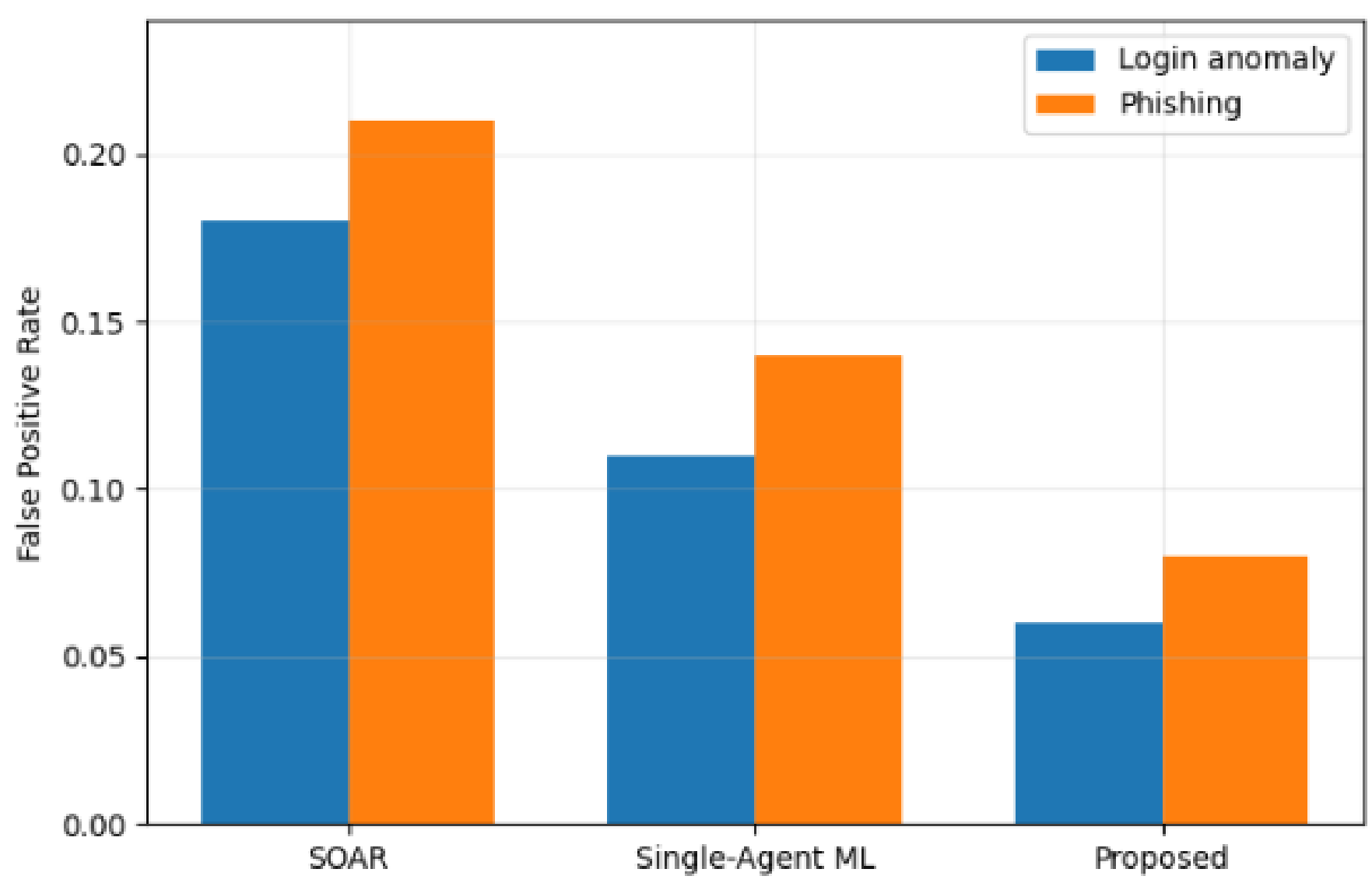

Figure 2 shows the false positive rate across login anomaly and phishing scenarios. The proposed framework achieves consistently lower false positive rates (approximately 0.06–0.08) compared to both the SOAR baseline and the single-agent model.

This reduction reflects the effect of meta-cognitive decision gating, which prevents premature or incorrect automated actions under uncertainty.

**Decision Behavior**

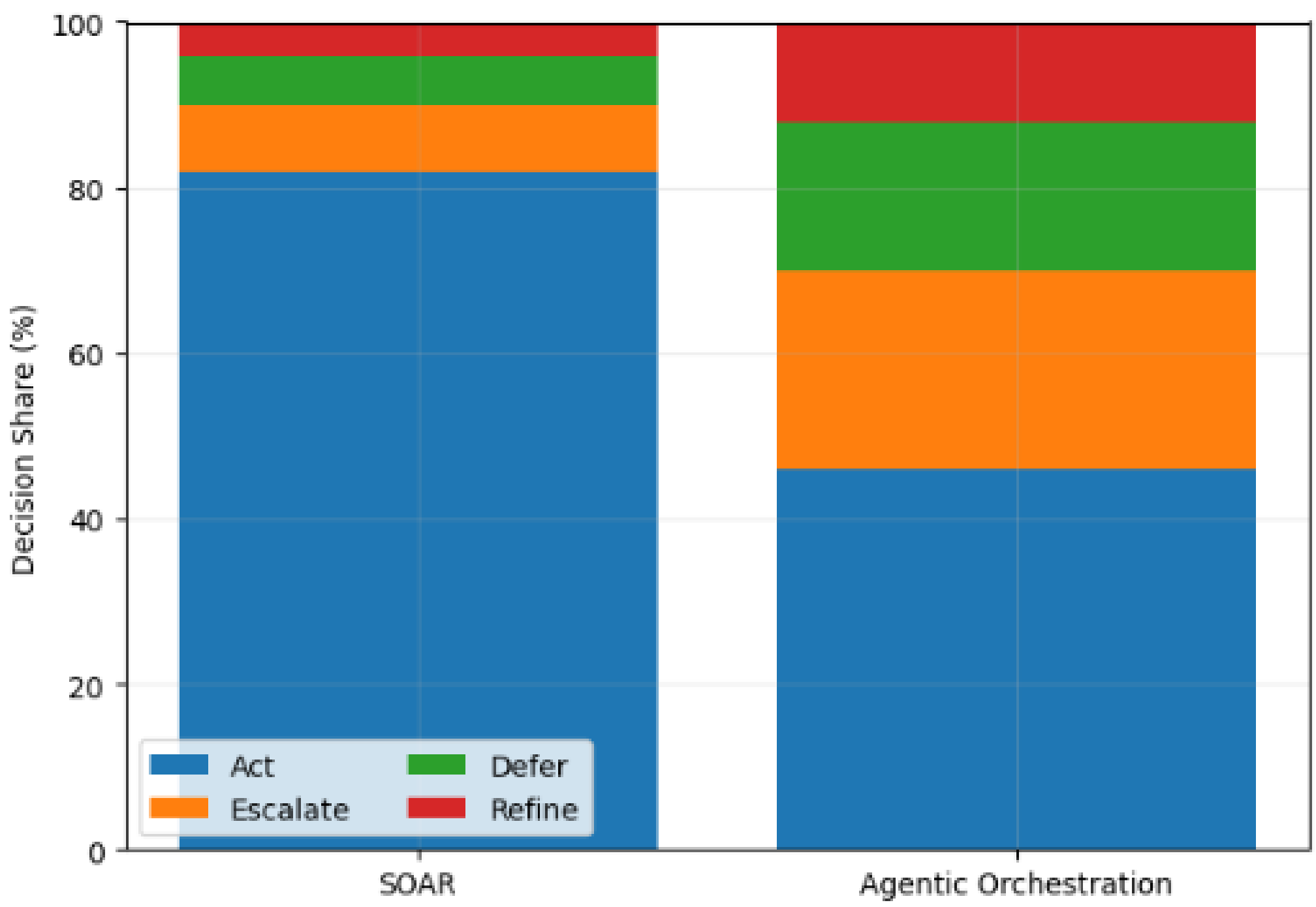

Figure 3 illustrates the distribution of decision outcomes across system configurations. The deterministic SOAR pipeline overwhelmingly favors immediate action (approximately 82%), whereas the proposed framework distributes decisions across action, escalation, deferral, and refinement.

This balanced distribution demonstrates a more adaptive and uncertainty-aware decision strategy aligned with human analyst behavior.

**Confidence Calibration**

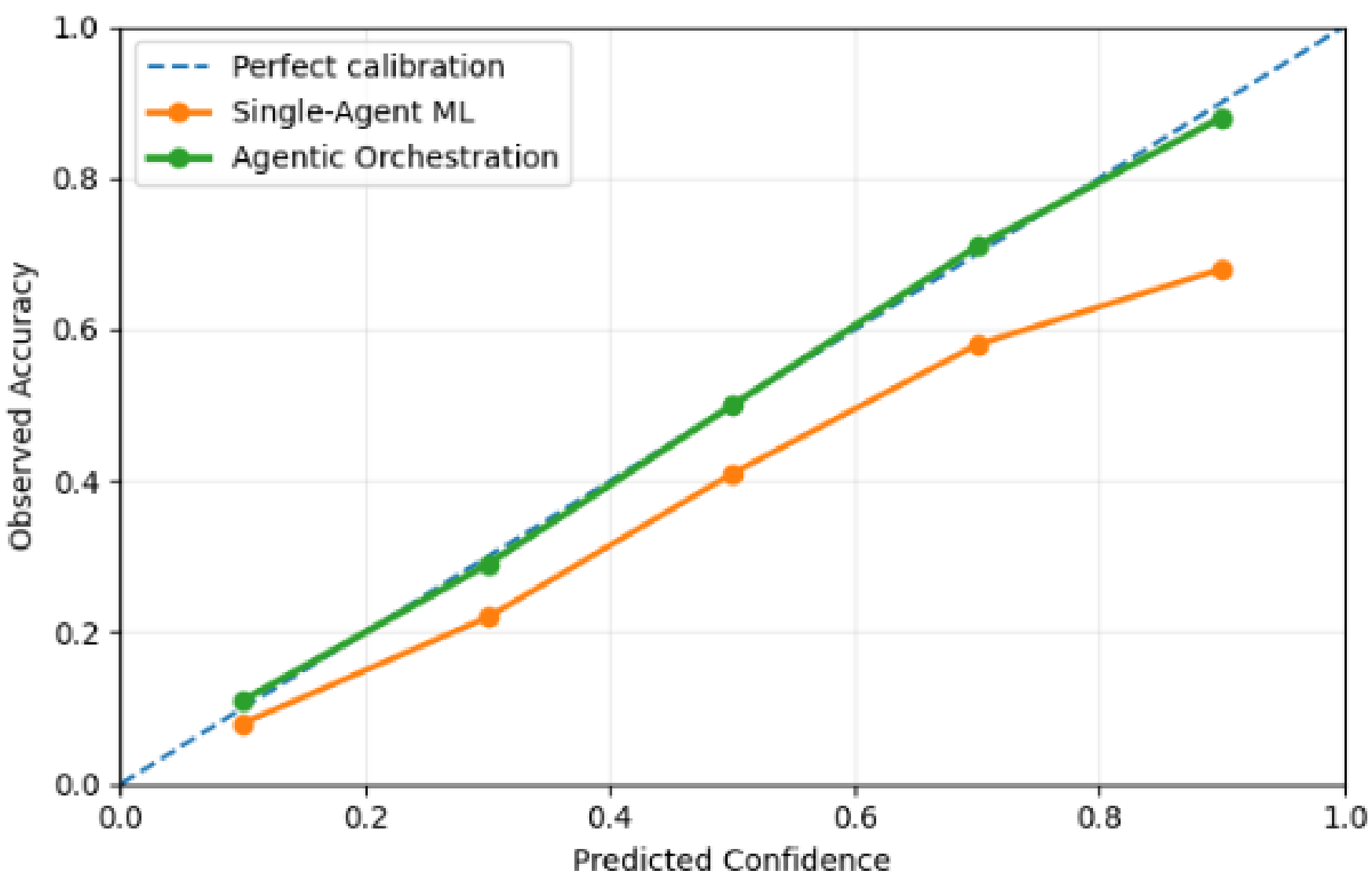

Figure 4 shows calibration curves comparing predicted confidence to observed accuracy. The proposed framework produces confidence estimates that closely align with actual accuracy, approaching ideal calibration, while the single-agent model exhibits systematic overconfidence.

This indicates improved reliability and interpretability of decision outputs.

### Robustness to Adversarial Perturbations

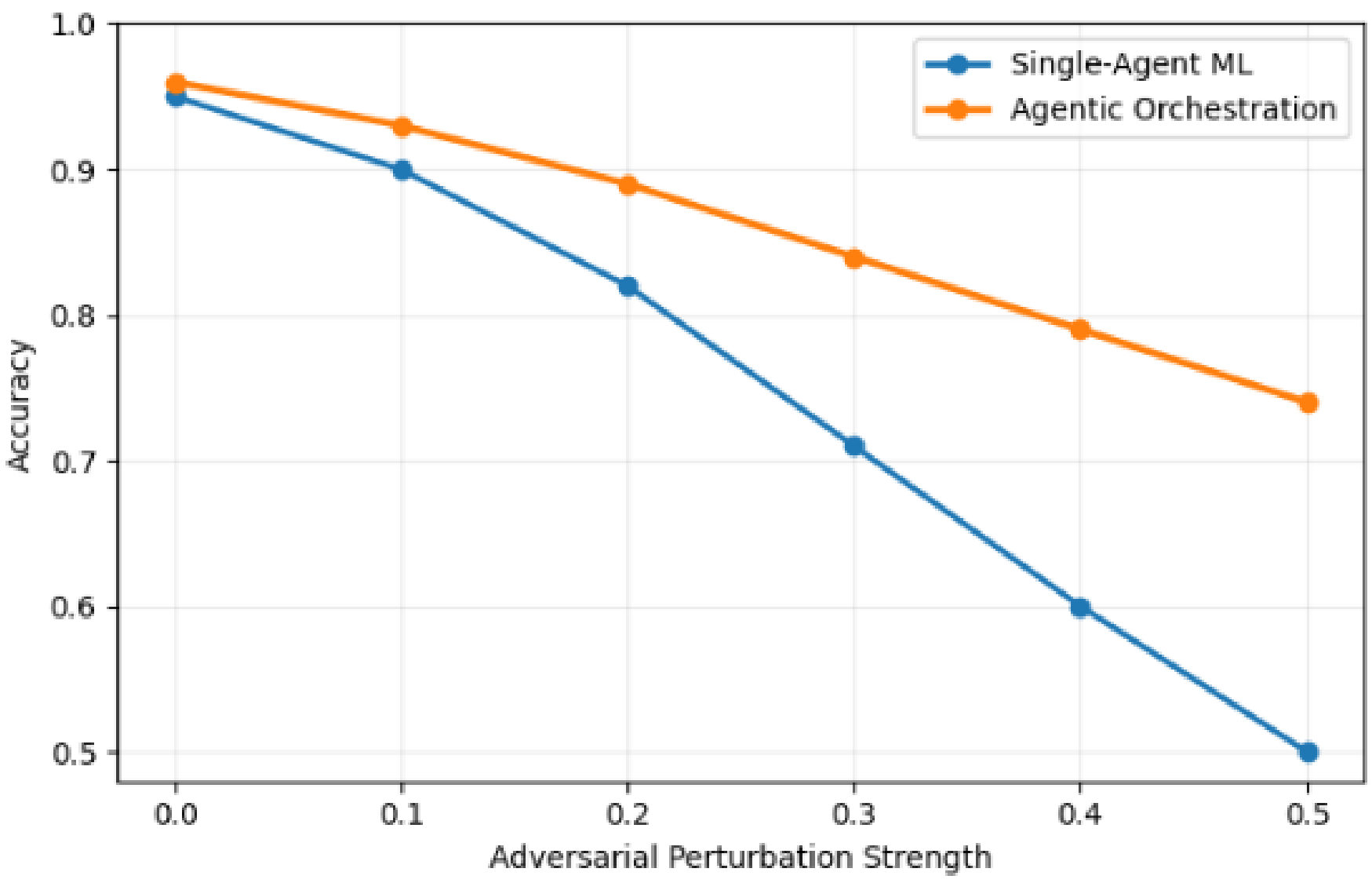

Figure 5 evaluates performance under increasing adversarial perturbation strength. The single-agent model shows substantial degradation, whereas the proposed framework degrades more gradually, maintaining higher accuracy under strong perturbations.

This demonstrates improved resilience to adversarial manipulation.

### Ablation Analysis

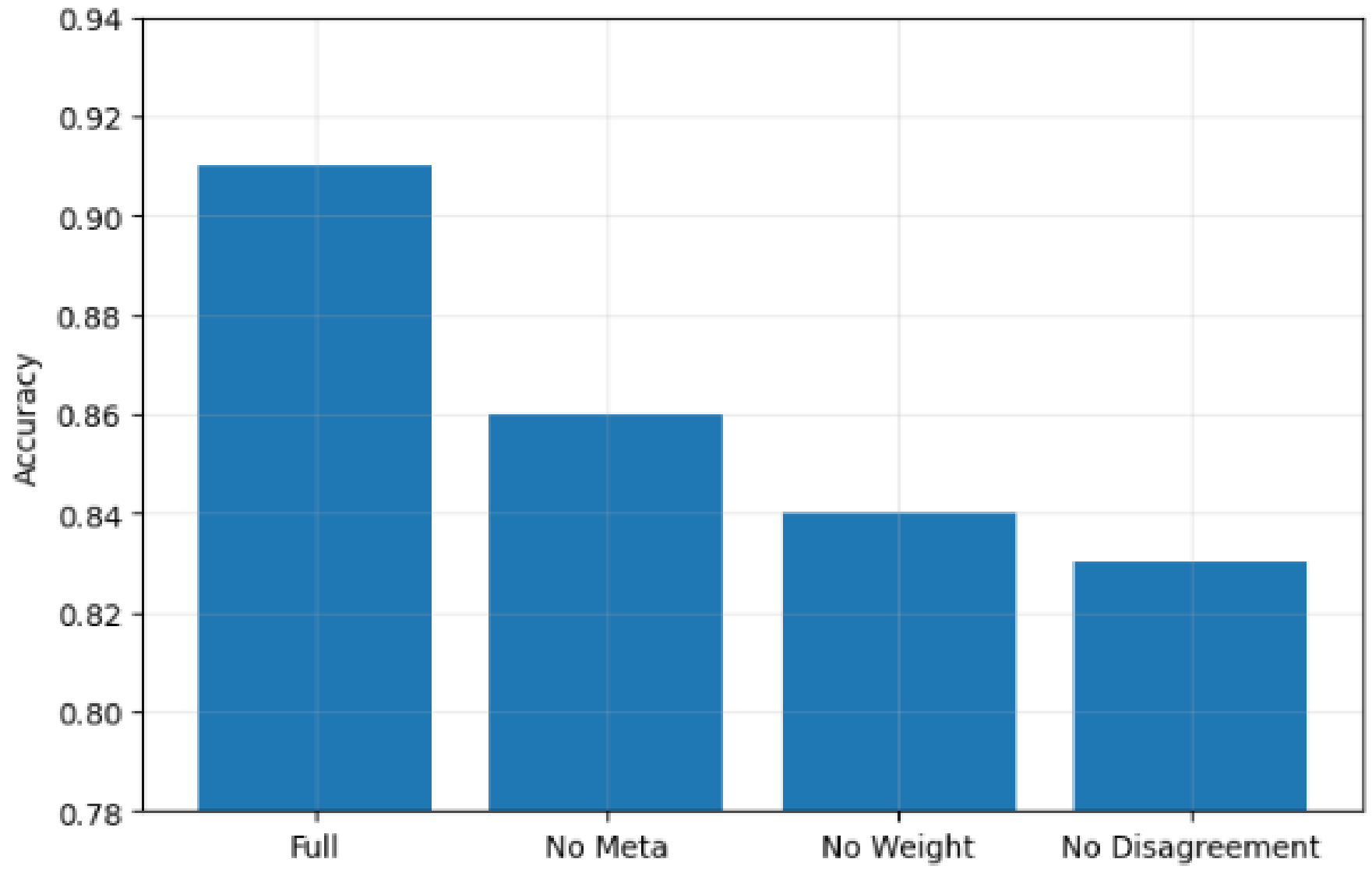

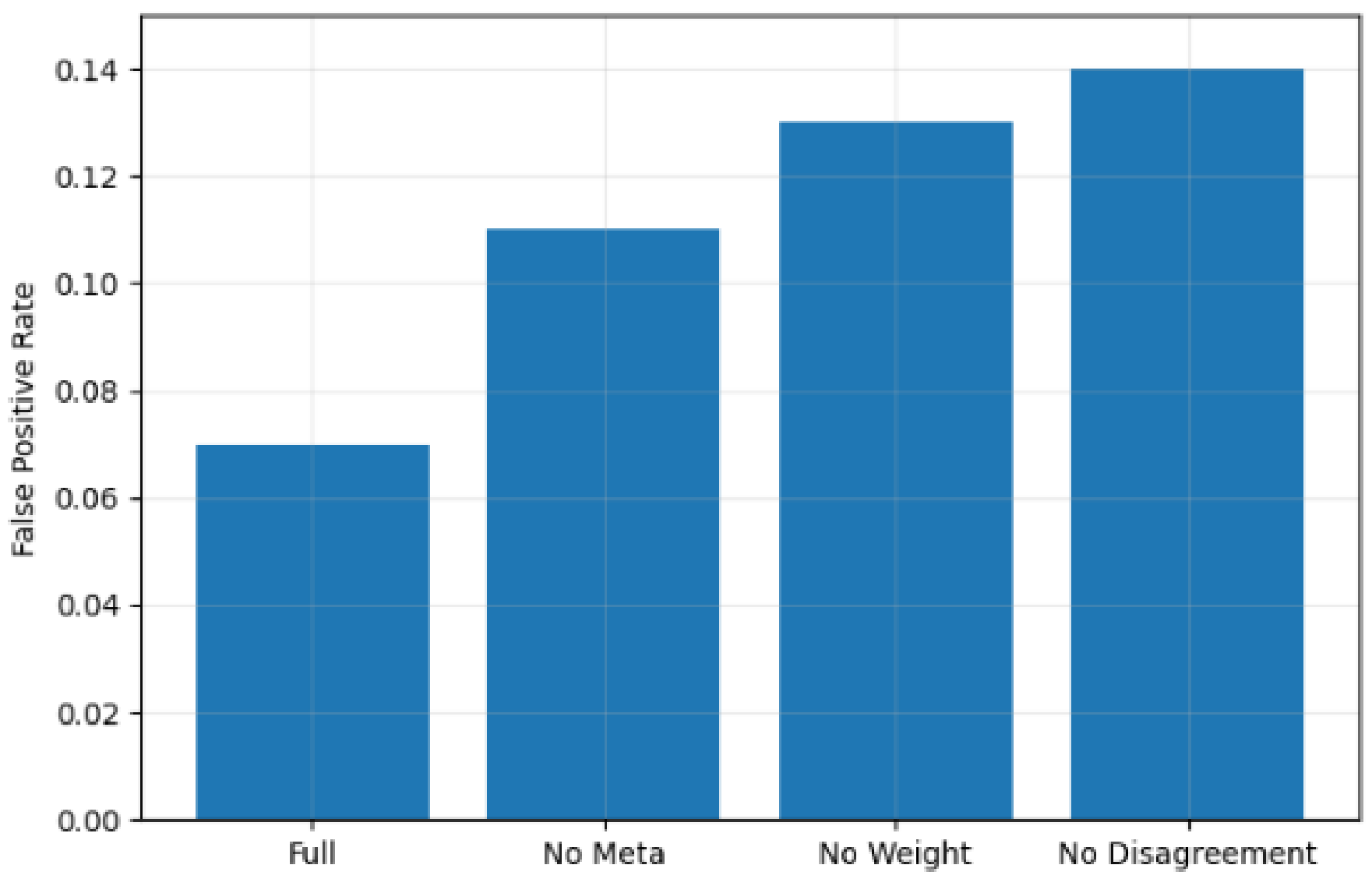

Figures 6A–6B show the impact of removing key components of the framework. Eliminating meta-cognitive judgement, confidence weighting, or disagreement modeling leads to reduced accuracy and increased false positive rates.

These results confirm that each component contributes to overall system performance and decision reliability.

## 5.7 Discussion of Results

The results demonstrate that incorporating:

- probabilistic belief aggregation
- meta-cognitive decision gating
- and disagreement-aware reasoning

significantly improves decision-making in cybersecurity environments.

Importantly, the system avoids overconfident and premature actions, instead adapting its behavior based on the **quality and coherence of available evidence**.

This aligns with real-world SOC workflows, where uncertainty and ambiguity are intrinsic rather than exceptional.

# 6. Discussion

The proposed framework reframes cybersecurity from a pipeline-oriented automation problem to a distributed cognitive process involving artificial and human agents. This shift has important implications for both system design and operational practice.

First, the results suggest that many limitations of current cybersecurity systems are not primarily computational, but cognitive. Contemporary systems are highly effective at detecting patterns, yet they often struggle to support reliable decision-making under uncertainty. Errors frequently arise not from a lack of detection capability, but from misalignment between signals, context, explanation, and action. By explicitly modeling meta-cognitive judgement, the proposed framework addresses this gap by regulating when and how actions should be taken, rather than simply improving what is detected.

Second, the introduction of meta-cognitive judgement as a first-class architectural element enables a form of **proportional and reflective autonomy**. Instead of treating autonomy as a fixed system property, the framework dynamically calibrates the level of automation based on uncertainty, disagreement, and contextual constraints. This aligns with emerging responsible AI principles, which emphasize meaningful human control, accountability, and transparency in high-stakes environments.

Third, the results highlight the importance of uncertainty-aware decision strategies in adversarial settings. The ability to defer, escalate, or refine decisions—rather than forcing immediate action—represents a significant departure from deterministic pipeline architectures. This behavior more closely reflects human analytical reasoning in Security Operations Centers (SOCs), where ambiguity and incomplete information are intrinsic to the task.

At the same time, the proposed approach introduces new challenges. Agentic orchestration and meta-cognitive judgement require additional computational resources and more complex system design. Their effectiveness depends on the quality and diversity of underlying agents, as well as on the calibration of uncertainty and confidence estimates. Moreover, deploying such systems in real-world environments raises organizational and socio-technical questions, including trust in automated judgement, responsibility allocation, and integration with existing workflows.

The exploratory use of student design projects provides preliminary insight into the cognitive accessibility of the framework. The relative ease with which participants adopted concepts such as decision readiness and autonomy regulation suggests that the meta-cognitive model aligns with intuitive human reasoning. However, the variability observed across implementations indicates the need for clearer design patterns, standardized interfaces, and supporting tools to facilitate practical adoption.

Finally, it is important to note that the benefits of meta-cognitive orchestration are context-dependent. In environments where actions are fully deterministic or easily reversible, the additional overhead of meta-cognitive reasoning may not be justified. However, in adversarial, high-risk, or high-uncertainty settings—such as cybersecurity operations—the ability to regulate decision-making processes becomes critical.

Overall, the findings suggest that advancing cybersecurity requires not only better detection models, but systems capable of reasoning about their own knowledge, uncertainty, and decision authority.

## 7. Conclusion

This paper proposed a conceptual re-framing of cybersecurity orchestration as an agentic, multi-agent cognitive system. By making explicit the roles of interpretation, explanation, governance, and meta-cognitive judgement, the framework moves beyond pipeline-centric architectures toward reflective and accountable autonomy.

The introduction of a meta-cognitive judgement function enables cybersecurity systems to reason about decision readiness and regulate their own autonomy under uncertainty, aligning technical capabilities with responsible AI principles and human oversight requirements. Rather than optimizing isolated detection or response tasks, the proposed approach emphasizes governable decision-making in adversarial and uncertain environments.

We argue that this perspective provides a foundation for next-generation AI-enabled cybersecurity systems and opens new directions for research at the intersection of AI systems design, cybersecurity operations, and socio-technical governance.

## 8. Bibliography

1. Sommer, R.; Paxson, V. Outside the Closed World: On Using Machine Learning for Network Intrusion Detection. *Proc. IEEE Symp. Security and Privacy (SP)* **2010**, 305–316. https://doi.org/10.1109/SP.2010.25.

2. Apruzzese, G.; Colajanni, M.; Ferretti, L.; Guido, A.; Marchetti, M. On the Effectiveness of Machine and Deep Learning for Cyber Security. In *Proc. 10th Int. Conf. on Cyber Conflict (CyCon)*; **2018**.

3. Wooldridge, M. *An Introduction to MultiAgent Systems*, 2nd ed.; Wiley: Chichester, UK, **2009**.

4. Hutchins, E. *Cognition in the Wild*; MIT Press: Cambridge, MA, USA, **1995**. https://doi.org/10.7551/mitpress/1881.001.0001.

5. Hollnagel, E.; Woods, D.D. *Joint Cognitive Systems: Foundations of Cognitive Systems Engineering*; CRC Press/Taylor & Francis: Boca Raton, FL, USA, **2005**.

6. Suchman, L. *Human–Machine Reconfigurations: Plans and Situated Actions*, 2nd ed.; Cambridge University Press: Cambridge, UK, **2012**. https://doi.org/10.1017/CBO9780511808418.

7. Miller, T. Explanation in Artificial Intelligence: Insights from the Social Sciences. *Artif. Intell.* **2019**, *267*, 1–38.

8. Doshi-Velez, F.; Kim, B. Towards a Rigorous Science of Interpretable Machine Learning. *arXiv* **2017**, arXiv:1702.08608.

9. Bender, E.M.; Gebru, T.; McMillan-Major, A.; Shmitchell, S. On the Dangers of Stochastic Parrots: Can Language Models Be Too Big? In *Proc. ACM Conf. on Fairness, Accountability, and Transparency (FAccT '21)*; **2021**; pp. 610–623. https://doi.org/10.1145/3442188.3445922.

10. National Institute of Standards and Technology (NIST). *Artificial Intelligence Risk Management Framework (AI RMF 1.0)*; NIST: Gaithersburg, MD, USA, **2023**. https://doi.org/10.6028/NIST.AI.100-1.

11. National Institute of Standards and Technology (NIST). *Artificial Intelligence Risk Management Framework: Generative Artificial Intelligence Profile (NIST AI 600-1)*; NIST "NVLAP": **2024**.

12. European Union. Regulation (EU) 2024/1689 of the European Parliament and of the Council of 13 June 2024 Laying Down Harmonised Rules on Artificial Intelligence (Artificial Intelligence Act). *OJ L* **2024**, 2024/1689.

13. Nelson, T.O.; Narens, L. Metamemory: A Theoretical Framework and New Findings. In Psychology of Learning and Motivation; Bower, G.H., Ed.; Academic Press: San Diego, CA, USA, 1990; Volume 26, pp. 125–173.

14. Cox, M.T. Metacognition in Computation: A Selected Research Review. Artificial Intelligence 2005, 169, 104–141. https://doi.org/10.1016/j.artint.2005.10.009

15. Floridi, L.; Cowls, J.; Beltrametti, M.; Chatila, R.; Chazerand, P.; Dignum, V.; Luetge, C.; Madelin, R.; Pagallo, U.; Rossi, F.; Schafer, B.; Valcke, P.; Vayena, E. AI4People—An Ethical Framework for a Good AI Society. Minds Mach. 2018, 28, 689–707. https://doi.org/10.1007/s11023-018-9482-5

16. Rafy, A.; Ahmed, M. Artificial Intelligence in Cybersecurity: A Comprehensive Multidomain Review of Techniques, Applications, Challenges, and Future Directions. *Electronics* **2023**, *12*, 4040. https://doi.org/10.3390/electronics12194040

17. Vinay, V. *The Evolution of Agentic AI in Cybersecurity: From Single LLM Reasoners to Multi-Agent Systems and Autonomous Pipelines.* arXiv **2025**, arXiv:2512.06659.

18. Jestus Lazer, S.; Aryal, K.; Gupta, M.; Bertino, E. *A Survey of Agentic AI and Cybersecurity: Challenges, Opportunities and Use-Case Prototypes.* arXiv **2026**, arXiv:2601.05293.

19. Arora, S.; Hastings, J. *Securing Agentic AI Systems — A Multilayer Security Framework.* arXiv **2025**, arXiv:2512.18043.

20. Gosmar, D.; Dahl, D.A. *Sentinel Agents for Secure and Trustworthy Agentic AI in Multi-Agent Systems.* arXiv **2025**, arXiv:2509.14956.

21. Wei, H.; Shakarian, P.; Lebiere, C.; Draper, B.; Krishnaswamy, N.; Sreedharan, S. *Metacognitive AI: Framework and the Case for a Neurosymbolic Approach.* In *Proc. Metacognitive AI Workshop*; Springer, 2024.